\documentclass[aps,prl,showpacs,superscriptaddress,nofootinbib,groupedaddress]{revtex4} 

\usepackage{color}
\usepackage{graphicx}
\usepackage{amsmath}
\usepackage{amssymb}
\usepackage[colorlinks=true,citecolor=darkred,urlcolor=darkred, pdfborder={0 0 0}]{hyperref}
\definecolor{darkred}{rgb}{0.6,0,0}

\definecolor{linkcolor}{rgb}{0,0,0.5}

\def\gsim{\raise0.3ex\hbox{$\;>$\kern-0.75em\raise-1.1ex\hbox{$\sim\;$}}}
\def\lsim{\raise0.3ex\hbox{$\;<$\kern-0.75em\raise-1.1ex\hbox{$\sim\;$}}}

\def\beqn#1{\begin{equation}\label{#1}}
\def\eeqn{\end{equation}}

\def\beqa#1{\begin{eqnarray}\label{#1}}
\def\eeqa{\end{eqnarray}}

\def\znbb {neutrinoless double beta decay }

\def\Z2{$\mathcal{Z_2}$}

\def\vev#1{\left\langle #1\right\rangle}

\newcommand{\sm}{{Standard Model }}
\newcommand{\AddrAHEP}{%
  AHEP Group, Institut de F\'{i}sica Corpuscular --
  C.S.I.C./Universitat de Val\`{e}ncia, Parc Cient\'ific de Paterna.\\
 C/ Catedr\'atico Jos\'e Beltr\'an, 2 E-46980 Paterna (Valencia) - SPAIN}
 \def\one{\ensuremath{\mathbf{1}}}
 \def\three{\ensuremath{\mathbf{3}}}
 \def\threeS{\ensuremath{\mathbf{3^*}}}	

\def\lnv{lepton number violating }
\def\SM{$\mathrm{SU(3)_c \otimes SU(2)_L \otimes U(1)_Y}$ }
\def\TrTrOne{ $\mathrm{SU(3)_c \otimes SU(3)_L \otimes U(1)_X}$ }

\begin{document}

\title{Dynamical seesaw mechanism for Dirac neutrinos}
\author{Jos\'e W.F. Valle}
\email{valle@ific.uv.es}
\affiliation{\AddrAHEP}
\author{C.A. Vaquera-Araujo}
 \email{vaquera@ific.uv.es}
\affiliation{\AddrAHEP}
\date{\today}

\pacs{14.60.Pq, 12.60.Cn, 14.60.St}


\begin{abstract}
\noindent

So far we have not been able to establish that, as theoretically
expected, neutrinos are their own anti-particles. Here we propose a
dynamical way to account for the Dirac nature of neutrinos and the
smallness of their mass in terms of a new variant of the seesaw
paradigm in which the energy scale of neutrino mass generation could
be accessible to the current LHC experiments.

\end{abstract}

\maketitle

\section{Introduction}

Although neutrinos are not proven experimentally to be Dirac or
Majorana fermions, the leading theoretical expectation is that they are
Majorana. Indeed there is a widespread paradigm that ascribes the
smallness of neutrino masses relative to the other \sm fermion masses
to their charge neutrality.
This is naturally incorporated by assuming that neutrinos acquire
Majorana masses from a \lnv operator, such as Weinberg's dimension
five operator or similar higher order ones. Indeed conventional
type-I~\cite{gell-mann:1980vs,yanagida:1979,mohapatra:1980ia,Schechter:1980gr,Schechter:1981cv}
or type-II~\cite{Schechter:1980gr,Schechter:1981cv,Lazarides:1980nt}
formulations of the seesaw mechanism lead typically to Majorana
neutrinos, irrespective of whether the seesaw is realized at high or
at low mass scale, in the spirit of the models considered
in~\cite{Boucenna:2014zba}.
Until the observation of \znbb~\cite{klapdor-kleingrothaus:2004wj}
becomes unambiguously confirmed~\cite{Barabash:2011fn} the possibility
remains that neutrinos can be Dirac particles after all.

The theoretical challenge to account for this possibility is then
twofold: i) to predict Dirac neutrinos, and ii), to understand
dynamically their small mass. Regarding the first we need to use extra
symmetries beyond \SM gauge symmetry, otherwise massive neutrinos are
generally expected to be Majorana
particles~\cite{Schechter:1980gr}. To this end, within the standard
\SM electroweak gauge structure, one may impose a conserved lepton
number, so as to obtain Dirac neutrinos. If the lepton number
assignment is non-standard one may obtain naturally light Dirac
neutrinos from calculable radiative
corrections~\cite{peltoniemi:1993ss}.
Likewise, one may consider schemes based on flavor symmetries, as
suggested in~\cite{Aranda:2013gga}. Unfortunately in the simplest
realization of this idea the smallness of neutrino mass is put in by
hand. Another approach would be to appeal to the existence of extra
dimensions~\cite{arkani-hamed:1998vp,Chen:2015jta}.

Alternatively one may extend the gauge group itself, so as to (at
least partially) include the lepton number symmetry, for example, by
using the extended \TrTrOne gauge structure~\cite{Singer:1980sw} which
predicts the number of fermion generations to match the number of
colors.
Although modern \TrTrOne schemes have Majorana neutrinos, with either
radiatively induced~\cite{Boucenna:2014ela,Boucenna:2014dia} or
seesaw--type masses~\cite{Boucenna:2015raa,Vien:2016tmh}, the simplest
original formulation predicts Dirac neutrino
masses~\cite{valle:1983dk}. Apart from not being able to account for
current oscillation data, that original formulation did not address
the question of how to account for the observed smallness of neutrino
mass.

In this letter we focus on the possibility of having naturally light
Dirac neutrinos with masses induced\textit{ a la seesaw}. We adopt the
\TrTrOne gauge structure because of its unique features, with respect
to other electroweak extensions based, for example, on left-right
symmetry.
Indeed, our new variant of the seesaw mechanism for Dirac neutrinos
makes use of the peculiar features of the \TrTrOne based models. The
resulting scheme provides a Dirac seesaw alternative to the approaches
considered in~\cite{Boucenna:2014ela,Boucenna:2014dia}
and~\cite{Boucenna:2015raa,Vien:2016tmh}.

\section{the Model}
\label{Model}

Our starting point is a variant of the \TrTrOne gauge framework with
the anomaly--free matter content given in Tab.~(\ref{tab:content}).
Notice that one has the same set of ``gauge--charged'' fields as
in~\cite{Boucenna:2014ela}, for example the left-handed leptons
transform as
\begin{equation}
\psi_L^\ell=\left(
\begin{array}{c}
\ell^-\\
\nu_\ell\\
N_\ell
\end{array} \right)_L  \,,
\end{equation}
with $\ell=1,2,3 \equiv e,\,\mu,\,\tau$.  In contrast to what is
assumed in~\cite{Boucenna:2014ela}, however, we add more gauge singlet
leptons and we change the transformation properties of two of the
scalar multiplets under $ \mathcal{L}$, the ungauged piece of the
lepton number symmetry.
Indeed, in this model the electric charge can be written in terms of
the $\mathrm{U(1)_X}$ generator $X$ and the diagonal generators of the
$\mathrm{SU(3)_L}$, whereas lepton number has a gauge component as
well as a complementary global one:
\beqa{eq:QL}
Q&=&T_3+\frac{1}{\sqrt{3}}T_8+X \, , \\
L&=&\frac{4}{\sqrt{3}} T_8+\mathcal{L}  \, .
\eeqa
\begin{table}[!h]
\begin{center}
\begin{tabular}{|c||c|c|c||c|c|c|c||c|c|c|}
\hline
 & $\psi_L^\ell$ & $\ell_R$ & $S_R^{\ell}\,,\tilde{S}_R^{\ell}$ & $Q_L^{1,2}$  & $Q_L^3$ & $\hat{u}_{R}$ & $\hat{d}_{R}$  & $\phi_0$ & $\phi_1$ & $\phi_2$ \\
\hline
\hline
$\mathrm{SU(3)_c}$ & \one &\one &\one &\three &\three &\three &\three &\one&\one &\one \\
\hline
$\mathrm{SU(3)_L}$ & \threeS & \one & \one &\three & \threeS& \one & \one & \threeS & \threeS& \threeS    \\
\hline
$\mathrm{U(1)_X}$ & $-\frac{1}{3}$ & $-1$ &  $0$ & $0$ & $+\frac{1}{3}$ & $+\frac{2}{3}$ & $-\frac{1}{3}$ & $+\frac{2}{3}$& $-\frac{1}{3}$& $-\frac{1}{3}$\\ 
\hline
$ \mathcal{L} $ & $-\frac{1}{3}$ & $-1$ &  $1$ & $-\frac{2}{3}$ & $+\frac{2}{3}$ & $0$ & $0$ &$+\frac{2}{3}$ & $-\frac{4}{3}$&$-\frac{4}{3}$\\ 
\hline
 $ \mathbb{Z}_3^{\rm aux} $ & $\omega$ & $\omega$ &  $\omega$ & $\omega^2$ & $\omega^2$ & $\omega^2$ & $\omega^2$ & $1$& $1$& $1$\\ 
\hline
\end{tabular}\caption{Matter content of the model, where
  $\hat{u}_R\equiv{(u_R,c_R,t_R,U_R)}$ and $\hat{d}_R\equiv{(d_R,s_R,b_R,D_R,D'_R)}$.} 
\label{tab:content}
\end{center}
\end{table}

Let us now explain the new features of the model in relation to
previous variants of the  \TrTrOne electroweak model.

\emph{Change in the matter sector}: In addition to the two-component
neutral fermions $N_L$ needed to fill up the $\mathrm{SU(3)_L}$
anti-triplets, in our new model we introduce \emph{two} sequential sets of $
\mathcal{L}$--carrying gauge singlet leptons denoted as
$S_R$, $\tilde{S}_R$.

\emph{Change in the scalar sector}: As before, three scalar
anti-triplets $\phi_{0}\sim ({\bf 3}^*,+2/3)$ and $\phi_{1,2} \sim ({\bf
  3}^*,-1/3)$ are responsible for the spontaneous breakdown of the
extended electroweak $\mathrm{SU(3)_L}$ gauge symmetry. However, in
contrast to the formulation presented in ~\cite{Boucenna:2014ela}, in
the present framework the scalar triplets $\phi_{1}$ and $\phi_{2}$
have the same $\mathcal{L}$ charge. Following the notation of
\cite{valle:1983dk} the most general pattern for the vacuum
expectation values (VEVs) of the fields is
\begin{equation}
\label{scalar3plets}
\vev{\phi_0}=\frac{1}{\sqrt{2}}\left(
\begin{array}{c}
k_0\\
0\\
0
\end{array}
\right),\qquad
\vev{\phi_1}=\frac{1}{\sqrt{2}}\left(
\begin{array}{c}
0\\
k_1\\
n_1
\end{array}
\right),\qquad
\vev{\phi_2}=\frac{1}{\sqrt{2}}\left(
\begin{array}{c}
0\\
k_2\\
n_2
\end{array}
\right)\,,
\end{equation}
where the isosinglet VEVs $n_1$ and $n_2$ characterize the
$\mathrm{SU(3)_L}$ breaking scale.  Correspondingly $k_0$, $k_1$ and
$k_2$ are the VEVs of the $\mathrm{SU(2)_L \subset SU(3)_L}$ doublets,
so we expect that $k_0\,, k_1\,,k_2\ll n_1\,,n_2$.\\[-.2cm]

In the fermion sector, the relevant Yukawa terms invariant under the
gauge symmetry and the auxiliary $\mathbb{Z}_3^{\rm aux}$ are
\begin{eqnarray}\label{lagY}
 -\mathcal{L}_{\rm f} &=& \,
 y^{\ell} \bar{\psi}_L \,l_R \,\phi_0 + y_1 \,\bar{\psi}_L \, S_{R} \,\phi_1 + \tilde{y}_{1} \,\bar{\psi}_L \, S_{R} \,\phi_2
 + y_2 \,\bar{\psi}_L \, \tilde{S}_{R} \,\phi_1 + \tilde{y}_{2} \,\bar{\psi}_L \, \tilde{S}_{R} \,\phi_2 \nonumber\\
 &+&  
y^{u}\, \bar{Q}_L^{1,2} \,\hat{u}_R \,\phi_0^*  +  
\hat{y}^{u}\, \bar{Q}_L^{3} \,\hat{u}_R \,\phi_1 +  
\tilde{y}^{u}\, \bar{Q}_L^{3} \,\hat{u}_R \,\phi_2 \nonumber \\
&+& 
y^{d} \,\bar{Q}_L^{3} \,\hat{d}_R \,\phi_0 + 
\hat{y}^{d} \, \bar{Q}_L^{1,2} \,\hat{d}_R\,\phi_1^* 
 +\tilde{y}^{d} \, \bar{Q}_L^{1,2} \,\hat{d}_R \,\phi_2^* 
 + \mathrm{h.c.}\,,
\end{eqnarray}
where contraction of the flavor indices is implicitly assumed.  An
important feature of the model is the fact that the $\mathcal{L}$
symmetry is preserved in the lepton sector. The role of the discrete
symmetry $\mathbb{Z}_3^{\rm aux}$ is to forbid the gauge invariant
$\,\psi_L^{ T} C^{-1} \psi_L \,\phi_0$ term in order to simplify the
following discussion.\\[-.2cm]

Concerning the symmetry breaking sector, the most general CP
conserving scalar potential compatible with the \TrTrOne gauge
symmetry as well as $\mathcal{L}$ invariance is given by
\begin{equation}
\begin{split}
V_{\mathcal{L}}=&\sum_i \left(\mu_i^2\left|\phi_i\right|^2+\lambda_i\left|\phi_i\right|^4\right)+\sum_{i\neq j} \lambda_{ij}\left|\phi_i\right|^2\left|\phi_j\right|^2+\sum_{i\neq j} \tilde{\lambda}_{ij}(\phi_i^{\dagger}\phi_j)( \phi_j^{\dagger}\phi_i)\\
&+\mu_s^2\left(\phi^{\dagger}_1\phi_2+\mathrm{h.c.}\right)+\sum_i \lambda'_i\left(\phi_1^{\dagger}\phi_2+\mathrm{h.c.}\right)\left|\phi_i\right|^2\\
&+\lambda\left[(\phi_1^{\dagger}\phi_2)(\phi_1^{\dagger}\phi_2)+\mathrm{h.c.}\right]+\tilde{\lambda}\left[(\phi_1^{\dagger}\phi_0)(\phi_0^{\dagger}\phi_2)+\mathrm{h.c.}\right]
\,,
\end{split}
\end{equation}
with $i=0,1,2$.  The key ingredient of our present construction is the
inclusion of a $\mathcal{L}$ violating piece in the scalar potential:
\begin{equation}
V_{\not {\mathcal{L}}}=f\left(\phi_0 \phi_1 \phi_2+\mathrm{h.c.}\right)\,,
\end{equation}
where the term $\phi_0 \phi_1 \phi_2$ is understood as the fully
antisymmetric product of the scalar field components. 
The mass dimension one parameter $f$ is expected to be small, in the
sense that the $\mathcal{L}$ symmetry is restored in the scalar sector
in the limit $f\to 0$~\footnote{The scalar potential in this limit has
  been studied in \cite{Giraldo:2011gd}.}. This symmetry can be
interpreted as a chiral symmetry which will control the massiveness of
neutrinos, as we show in the next section.

For simplicity, all VEVs and Yukawa coefficients are assumed to be
real in the following analysis.  From the minimization conditions of
the total potential $V=V_{\mathcal{L}}+V_{\not {\mathcal{L}}}$:
$\partial V/\partial k_0=\partial V/\partial n_1=\partial V/\partial
n_2=\partial V/\partial k_2=0$, the dimensionful parameters
$\mu_{0,1,2}$ and $f$ are determined as
\begin{equation}\label{min}
\begin{split}
\mu_0^2=&-\frac{1}{2}\left\{ 2\lambda_{0}k_{0}^2+\lambda_{01} 
   \left(k_{1}^2+n_{1}^2\right)+\lambda_{02}  \left(k_{2}^2+n_{2}^2\right)+\frac{\lambda'_{0} \left[k_{1}^2 \left(2
   k_{2}^2+n_{2}^2\right)+2 k_{1} k_{2} n_{1} n_{2}+n_{1}^2 \left(k_{2}^2+2
   n_{2}^2\right)\right]}{ k_{1} k_{2}+n_{1} n_{2}}\right\}\\
   &-\frac{ (k_{2} n_{1}-k_{1} n_{2})^2}{2
   k_{0}^2}\left[\tilde{\lambda}_{12}+2\lambda+\frac{2\mu_s^2+\lambda'_{1} \left(k_{1}^2+n_{1}^2\right) +\lambda'_{2} \left(k_{2}^2+n_{2}^2\right)}{
   k_{1} k_{2}+n_{1} n_{2}}\right]\,,\\
\mu_1^2=&-\frac{1}{2}\left\{\lambda_{01}k_{0}^2 +2\lambda_{1}
   \left(k_{1}^2+n_{1}^2\right)+(2\lambda  +\lambda_{12} 
   +\tilde{\lambda}_{12})  \left(k_{2}^2+n_{2}^2\right)+\frac{\left(k_{2}^2+n_{2}^2\right)\left[\lambda'_{0} k_{0}^2  +\lambda'_{2} \left(k_{2}^2+n_{2}^2\right)
   +2\mu_s^2 \right]}{k_{1} k_{2}+n_{1}
   n_{2}}\right\} \\
&  -\frac{\lambda'_{1} \left[k_{1}^2 \left(3
   k_{2}^2+n_{2}^2\right)+4 k_{1} k_{2} n_{1} n_{2}+n_{1}^2 \left(k_{2}^2+3
   n_{2}^2\right)\right]}{2 (k_{1} k_{2}+n_{1} n_{2})}   \,,\\
\mu_2^2=&-\frac{1}{2}\left\{\lambda_{02}k_{0}^2 +2\lambda_{2} \left(k_{2}^2+n_{2}^2\right)+(2\lambda + \lambda_{12} +
   \tilde{\lambda}_{12}) \left(k_{1}^2+n_{1}^2\right)+\frac{ \left(k_{1}^2+n_{1}^2\right)
   \left[\lambda'_{0}k_{0}^2 +\lambda'_{1}
   \left(k_{1}^2+n_{1}^2\right)+2\mu_s^2\right]}
   {k_{1} k_{2}+n_{1} n_{2}}\right\}\\
&-\frac{\lambda'_{2} \left[k_{1}^2 \left(3
   k_{2}^2+n_{2}^2\right)+4 k_{1} k_{2} n_{1} n_{2}+n_{1}^2 \left(k_{2}^2+3
   n_{2}^2\right)\right]}{2 (k_{1} k_{2}+n_{1} n_{2})}\,,\\
f=&(k_{1} n_{2}-k_{2} n_{1}) \left\{\frac{2 \lambda +\tilde{\lambda}_{12}}{\sqrt{2} k_{0}}+\frac{\lambda'_{0} k_{0}^2 
+ \lambda'_{1}(k_{1}^2+n_{1}^2)+\lambda'_{2} \left(k_{2}^2+n_{2}^2\right)+2 \mu_s^2}{\sqrt{2} k_{0}
   (k_{1} k_{2}+n_{1} n_{2})} \right\}   \,,  
\end{split}
\end{equation}
with the remaining condition $\partial V/\partial k_1=0$ automatically
satisfied. We conclude this section pointing out the interesting
dependence of $f$ on the VEV combination $(k_{1} n_{2}-k_{2}
n_{1})$. This means that we can interpret the last relation of
Eq.(\ref{min}) as a statement that a small parameter $f$ dynamically
induces a small non-zero value for $(k_{1} n_{2}-k_{2} n_{1})$.

\section{neutrino masses}
\label{sec:neutrino-msses}

After
spontaneous symmetry breaking, the Dirac neutrino mass matrix becomes
\begin{equation}
\begin{split}
-\mathcal{L}_{\rm mass}=\frac{1}{\sqrt{2}}\left(\begin{array}{cc}
\bar{\nu}_L & \bar{N}_L
\end{array}  \right) 
\left(\begin{array}{cc}
y_{1} k_1+\tilde{y}_{1} k_2 & y_{2} k_1+\tilde{y}_{2} k_2\\
y_{1} n_1+\tilde{y}_{1} n_2 & y_{2} n_1+\tilde{y}_{2} n_2\\
\end{array}\right)
\left(\begin{array}{c}
S_R\\
\tilde{S}_R
\end{array}  \right) + \mathrm{h.c.},
\end{split}
\end{equation}
where $y_{1,2}$ and $\tilde{y}_{1,2}$ are $3\times 3$ Yukawa matrices.
The light neutrino masses can be readily estimated in the one family
approximation, in which the neutrino mass matrix is diagonalized by a
bi-unitary transformation $\mathcal{M}_{\rm
  diag}=U_{\nu}^{\dagger}\mathcal{M}U_{S}$, with
\begin{equation}
\begin{split}
U_{\alpha}\approx\left(\begin{array}{cc}
\cos{\theta_\alpha} & \sin{\theta_\alpha}\\
-\sin{\theta_\alpha} & \cos{\theta_\alpha}
\end{array}  \right)\, ,\qquad 
\alpha=\nu,\, S\,,
\end{split}
\end{equation}
and
\begin{equation}
\begin{split}
\tan 2\theta_\nu&=-\frac{2 \left[k_{1}n_{1} \left(y_{1}^2+y_{2}^2\right)+(k_{1}n_{2}+ k_{2} n_{1})(\tilde{y}_{1} y_{1}+\tilde{y}_{2}
   y_{2})+k_{2} n_{2}
   \left(\tilde{y}_{1}^2+\tilde{y}_{2}^2\right)\right]}{ (k_{1}^2-n_{1}^2) \left(y_{1}^2+y_{2}^2\right)+2 (k_{1} k_{2}-n_{1} n_{2}) (\tilde{y}_{1} y_{1}+\tilde{y}_{2} y_{2})+(k_{2}^2-n_{2}^2)
   \left(\tilde{y}_{1}^2+\tilde{y}_{2}^2\right)}\,,\\
\tan 2\theta_S&=-\frac{2 \left[\left(k_{1}^2+n_{1}^2\right) y_{1} y_{2} +(k_{1} k_{2} + n_{1} n_{2})(\tilde{y}_{1} y_{2}+\tilde{y}_{2}
   y_{1})+(k_{2}^2 +n_{2}^2)
   \tilde{y}_{1} \tilde{y}_{2}\right]}{\left(k_{1}^2+n_{1}^2\right) (y_{1}^2-y_{2}^2)+2 (k_{1} k_{2}+n_{1}
   n_{2}) (\tilde{y}_{1} y_{1}-\tilde{y}_{2} y_{2})+(k_{2}^2+n_{2}^2) (\tilde{y}_{1}^2-\tilde{y}_{2}^2)}\,,
\end{split}
\end{equation}
yielding eigenstates with masses
\begin{equation}
\begin{split}
m_{\mp}&=\frac{1}{2}\sqrt{\left(A\mp\sqrt{A^2-4 B^2}\right)}\,,\\
A&=\left(k_{1}^2+n_{1}^2\right) \left(y_{1}^2+y_{2}^2\right)+2( k_{1} k_{2}+n_{1} n_{2} )(\tilde{y}_{1} y_{1}+\tilde{y}_{2}
   y_{2})+(k_{2}^2+n_{2}^2) \left(\tilde{y}_{1}^2+\tilde{y}_{2}^2\right)\,,\\
B&=(k_{1} n_{2}-k_{2} n_{1}) ( y_{1} \tilde{y}_{2}- y_{2}\tilde{y}_{1})  \,.
\end{split}
\end{equation}
Thus, up to corrections of $\mathcal{O}(B^2/A^{3/2})$, the resulting
masses for the light and heavy neutrinos are
\begin{equation}\label{mlh}
\begin{split}
m_{\rm light}&\approx\frac{ |(k_{1} n_{2}-k_{2} n_{1}) ( y_{1} \tilde{y}_{2}- y_{2}\tilde{y}_{1})| }{\sqrt{2\left[\left(k_{1}^2+n_{1}^2\right) \left(y_{1}^2+y_{2}^2\right)+2( k_{1} k_{2}+n_{1} n_{2} )(\tilde{y}_{1} y_{1}+\tilde{y}_{2}
   y_{2})+(k_{2}^2+n_{2}^2) \left(\tilde{y}_{1}^2+\tilde{y}_{2}^2\right)\right]}}\,,\\
m_{\rm heavy}&\approx\sqrt{\frac{1}{2}\left[\left(k_{1}^2+n_{1}^2\right) \left(y_{1}^2+y_{2}^2\right)+2( k_{1} k_{2}+n_{1} n_{2} )(\tilde{y}_{1} y_{1}+\tilde{y}_{2}
   y_{2})+(k_{2}^2+n_{2}^2) \left(\tilde{y}_{1}^2+\tilde{y}_{2}^2\right)\right]}\,.
\end{split}
\end{equation}

\begin{figure}[htb] \centering
   \includegraphics[width=.4\linewidth]{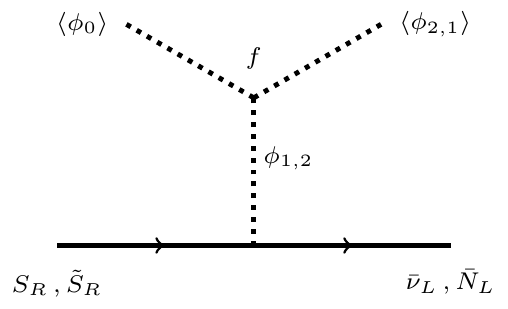}
   \caption{\label{fig:dss} Type--II--like dynamical seesaw mechanism
     for Dirac neutrino mass.}
\end{figure}
The mixing angle $\theta_\nu$ is small by virtue of the VEV hierarchy
$k_0\,, k_1\,,k_2\ll n_1\,,n_2$~\footnote{Notice that $\theta_S$
  becomes maximal in the limit $y_1\to y_2$, $\tilde{y}_1\to
  \tilde{y}_2$.}. For large $n_{1}, n_{2}$ VEVs one clearly obtains
the standard seesaw behavior, in which the small neutrino mass
emerges from the parameters governing the scale characterizing the
mass of the messenger particle, in this case a heavy scalar boson, see
Fig.~\ref{fig:dss}.

However Eq.(\ref{mlh}) contains further crucial information, namely
the fact that the light neutrino mass is determined by the scale $
(k_{1} n_{2}-k_{2} n_{1})$, the same combination of VEVs found in the
last relation in Eq.(\ref{min}).  The smallness of the neutrino mass
can then be understood as a consequence of the interplay between the
$f$ term in the scalar potential and $V_{\mathcal{L}}$. The presence
of a small $f$ parameter (quantifying the amount of $\mathcal{L}$
violation in the scalar sector) and large quartic couplings enforces a
nearly parallel dynamical alignment for the $\phi_2$ and $\phi_3$
VEVs, which in turn can lead to a tiny mass for the active neutrino
even without imposing a large hierarchy among the $k_{1,2}$ and
$n_{1,2}$ scales. 

Written in terms of $f$, and using the approximation $n_{1,2}\sim n\gg
k_{1,2}\sim k\,, k_0\,, |\mu_s|$, the mass of the light neutrino
simplifies to
\begin{equation}\label{ml2}
m_{\rm light}\approx \frac{  |f k_0 ( y_{1} \tilde{y}_{2}- y_{2}\tilde{y}_{1})|}{|\tilde{\lambda}_{12}+\lambda'_1+\lambda'_2+2\lambda | n \sqrt{ \left(y_{1}+\tilde{y}_{1}\right)^2+\left(y_{2}+\tilde{y}_{2}\right)^2}}\,.
\end{equation}
Assuming $|\tilde{\lambda}_{12}+\lambda'_1+\lambda'_2+2\lambda
|\sim\mathcal{O}(1)$ in the above equation, we can see that the
resulting mass is potentially suppressed by three different sources:
(i) the factor $k_0/n$; (ii) the small scale $f$ associated to the
$\mathcal{L}$--symmetry protection, and (iii) the determinant-like
Yukawa combination $(y_{1} \tilde{y}_{2}- y_{2}\tilde{y}_{1})$.  
We conclude this section with an illustrative example of how these
three sources can act in synergy: setting $f\sim \mathcal{O}(1)\,
\mathrm{keV}$, $k_0\sim\mathcal{O}(10^2)\, \mathrm{GeV}$,
$n\sim\mathcal{O}(10)\, \mathrm{TeV}$, and
$y_{1}=\tilde{y}_{2}=y+\delta$, $y_{2}=\tilde{y}_{1}=y-\delta$ with
$\delta\sim\mathcal{O}(10^{-2})$, a naturally small neutrino mass
$m_{\rm light}\sim \mathcal{O}(10^{-1})\, \mathrm{eV}$ is obtained
without the need to invoke superheavy physics.
The above estimate of light neutrino makes use of the one family
approximation. However, generalization to three families is
straightforward, using the perturbative block-diagonalization 
technique developed in \cite{Schechter:1981cv}.

\section{Discussion and conclusions}
\label{disc}
Summarizing, in this letter we have presented a novel mechanism for
Dirac neutrino mass generation in the context of \TrTrOne based
models. The light neutrino mass is induced at tree-level \textit{a la
  seesaw} as indicated in Fig.~\ref{fig:dss}. Its smallness is
guaranteed by three independent features:
\begin{enumerate}
\item The suppression imposed by the ratio of the $\mathrm{SU(2)_L}$
  and $\mathrm{SU(3)_L}$ breaking scales, which by itself can account
  for the small neutrino mass by invoking a large hierarchy among
  these scales as in the standard high-scale seesaw mechanism.
\item The dynamical alignment of the VEVs, induced by the presence of
  a small trilinear term in the scalar potential, which is ultimately
  related to the smallness of $\mathcal{L}$ violation encoded by the
  characteristic scale $f$.  The presence of this key ingredient makes
  it possible to have a low-scale realization of our seesaw mechanism,
  at an energy scale accessible to the current LHC experiments.
\item The additional suppression provided by the peculiar dependence
  of the neutrino mass on the Yukawa coefficients of the model, which
  favors a small neutrino mass in the case of a Yukawa alignment
  analogous to the one displayed by the VEVs. Potentially this might
  have a dynamical origin say, in string theories.
\end{enumerate} 
Before closing let us note that a low-scale seesaw realization in the
present context would bring in a very rich phenomenology with new
quarks, new gauge bosons as well as new scalars, all of them lurking
within the range that can be explored in flavor studies and the
LHC. Its detailed study lies beyond the scope of this letter. In
particular the model can fit the recent hint for a di-photon resonance
in a natural way~\cite{Boucenna:2015pav,Dong:2015dxw}.

\section*{Acknowledgements}
This work is supported by the Spanish grants FPA2014-58183-P,
Multidark CSD2009-00064, SEV-2014-0398 (MINECO) and
PROMETEOII/2014/084 (Generalitat Valenciana). C.A.V-A. acknowledges
support from CONACyT (Mexico), grant 251357.

\bibliography{d331}

\end{document}